\documentclass{article} 
\usepackage{hyperref}
\usepackage{graphicx}
\usepackage{amsmath}
\usepackage{url}
\usepackage{authblk}
\usepackage[margin=1in]{geometry}

\begin{document}

\title{ Combining exome and gene expression datasets in one graphical model of disease to empower the discovery of disease mechanisms}

\author[1,2]{Aziz M. Mezlini}
\author[2]{Fabio Fuligni}
\author[2]{Adam Shlien }
\author[1,2]{Anna Goldenberg\thanks{D.D@university.edu}}
\affil[1]{Department of Computer Science, University of Toronto}
\affil[2]{Hospital for Sick Children, Toronto}

\renewcommand\Authands{ and }

\newcommand{\fix}{\marginpar{FIX}}
\newcommand{\new}{\marginpar{NEW}}
\newcommand{\acomment}[1]{\emph{Anna: #1}}
\newcommand{\comment}[1]{}

\maketitle

\abstract{
Identifying genes associated with complex human diseases is one of the main challenges of human genetics and computational medicine. To answer this question, millions of genetic variants get screened to identify a few of importance. To increase the power of identifying genes associated with diseases and to account for other potential sources of protein function aberrations, we propose a novel factor-graph based model, where much of the biological knowledge is incorporated through factors and priors. Our extensive simulations show that our method has superior sensitivity and precision compared to variant-aggregating and differential expression methods. Our integrative approach was able to identify important genes in breast cancer, identifying genes that had coding aberrations in some patients and regulatory abnormalities in others, emphasizing the importance of data integration to explain the disease in a larger number of patients. 
}

\section{Introduction}

The majority of human diseases are complex, arising due to a multitude of factors. Identifying these factors is critical to understanding diseases and improving healthcare, yet it is a very difficult computational problem. Taking just one modality of data, such as the human genome, already presents computational difficulties. First, there are millions of common variants called Single Nucleotide Polymorphisms (SNPs). Computationally, these data can be represented as a matrix of variables (SNPs, usually taking on values of 0,1,2) measured for each patient and control for hundreds to thousands of individuals. The goal is to select features relevant to the disease, knowing that only a handful of features among these millions are actually relevant to the outcome (phenotype or disease). In the field of human genetics, this problem has been primarily addressed by univariate testing \cite{McCarthy2008}, where each variant is tested for a significant difference in distribution of values among patients vs healthy individuals. Due to a substantial multiple hypothesis correction problem, many studies have been underpowered to find any variants associated with the disease. While other, more complex methods, such as regularized regression \cite{Kim2009,Yang2009} were proposed to solve this problem, the biotechnological leap forward has presented new challenges.

In the last five years, sequencing has become a commonplace due to the lowering cost of the available biotechnology. While now a lot more variants are being covered, which means that it is more likely that the actual causal variant is among the sequenced ones, the computational problem has become more complex in many ways. First, the number of variants has increased, while the number of samples has decreased (due to the cost of sequencing). Second, now rare variants are available in addition to the previously available common variants. Rare variants may be present in only a handful of patients in the study and thus the standard approach to analyzing rare variants is to \emph{pool} them together, creating one meta variable per gene. These meta variables are then tested for association with the disease status in the same way as the common variants were. There are several such methods that are in broad use today including burden tests  such as CAST \cite{Morgenthaler2007} (collapsing all variants within genes). There are also methods testing the variance component such as C-alpha \cite{Neale2011} and SKAT \cite{Wu2011} which are more robust to the presence of neutral and protective variants.

The major problem with the above approaches is that they only work with coding variants, i.e. variants that are in the exome – protein-coding sequence. While those variants are indeed more likely to be harmful, disrupting the creation of a healthy protein, recent studies made it abundantly clear that the regulatory variation (e.g. genetic variants in regions between genes) play enormous role \cite{birney2007identification}. These variants are covered by the whole-genome sequencing (WGS) technology. However, WGS provides only a partial answer: it is currently too expensive to sequence a large number of patients, so the sample sizes remain too small compared to the number of variants. Thus, the most common way the WGS data is used is by restricting the analysis to a few novel variants (appearing just in that patient). The rest of the sequence is essentially wasted. However, even if there were enough patients with whole genome sequence available, this data would not be able to conclusively explain the abnormalities in the protein quantities, since there are many reasons for those changes, including epigenetic modifications that are not part of the DNA sequence. It is thus imperative to go beyond DNA to identify genes and proteins that are associated with the disease.

Outside of the field of genetics, there has been much work in identifying genes that are related to disease through differential gene expression DGE \cite{Anders2010}. These studies have again used a small number of patients and the approaches were mostly univariate (with a few exceptions through gene collections known as pathways or networks \cite{Jia2014}). Gene expression in itself is very noisy and hard to replicate, so few of the gene signatures identified through DGE have been adopted in the clinic.

 	The two data sources – coding variants and gene expression are mostly complementary, responsible for  different kinds of protein aberrations. We thus propose to combine these two data sources to a) improve the power of detecting genes associated with the disease; b) implicate proteins that have been affected in the population in a variety of ways, rather than solely through the DNA sequence. To this extent we propose a biologically motivated hierarchical factor graph model which efficiently combines these two sources of data. We use a variety of regularizing factors and prior knowledge, such as variant harmfulness and gene interactions as priors, to increase the likelihood of identifying the variants correctly.  To our knowledge, this is the first work that takes into account complementarity of exome and gene expression data sources in a principled way. It is also the first method that integrates external annotations, such as variant harmfulness and gene interaction information within the inference of the model. Our extensive simulations confirm that our method has superior sensitivity and precision compared to methods that combine rare variants. We have tested our approach in a large breast cancer dataset as a proof of concept and found that our method is able to identify important breast cancer genes. Interestingly, we find genes that have mutations in some patients and gene expression aberrations in other patients, indicating that our method is able to effectively explain the disease in more patients.

\section{Methods}

\begin{figure}[!h]
\centerline{\includegraphics[scale=0.35]{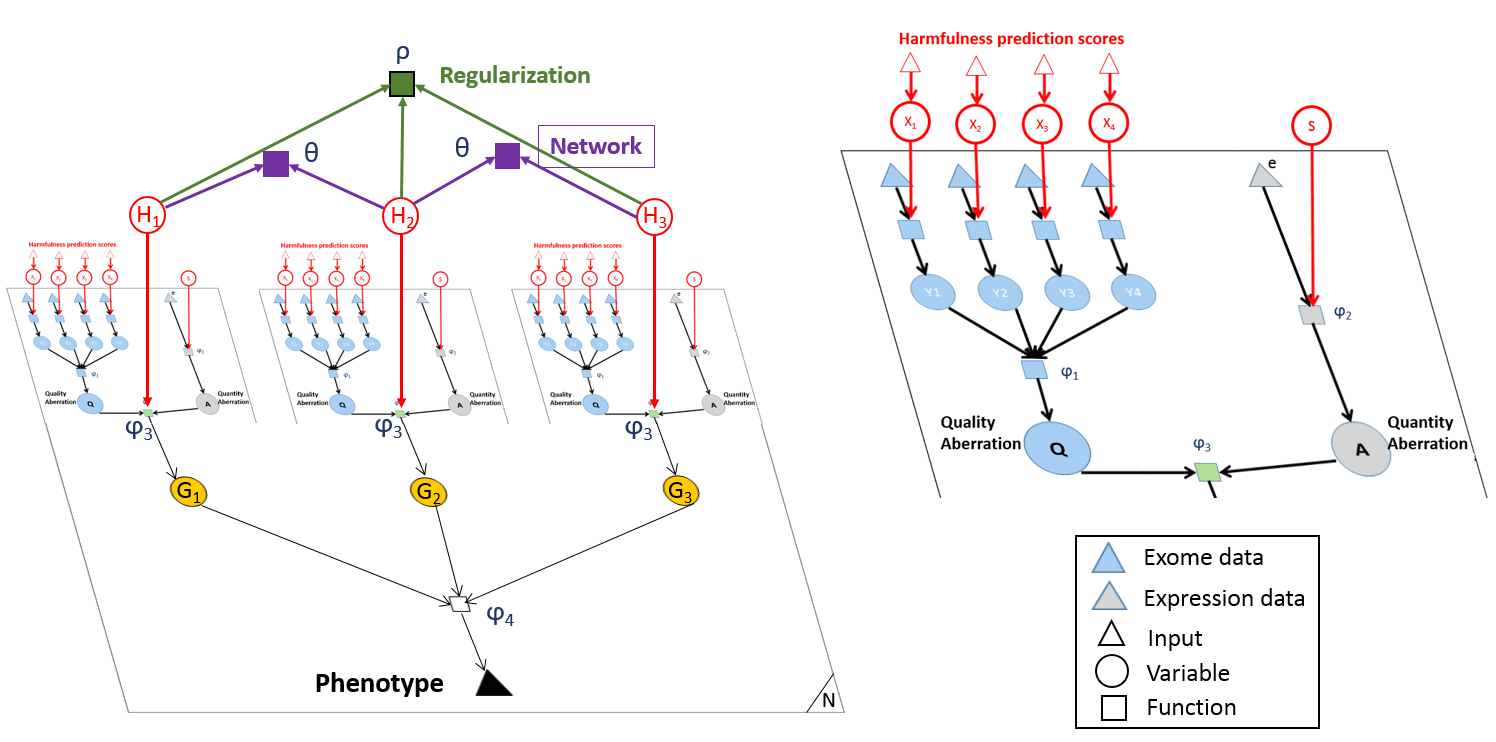}}
\caption{Graphical model for a genetic disease. All the variables, factors and inputs inside the plane are individual-specific and the whole plane should be replicated for every individual. The variables, factors and inputs outside the plane are not individual-specific. }
\label{fig:11}
\end{figure}

Our method jointly models the outcome (disease/no disease) and genotypes in a hierarchical graphical model. The predictor variables are the exome variants and mRNA expression levels for each person (patient or healthy individual). Figure \ref{fig:11} shows a factor graph corresponding to our model.

\paragraph{Modeling coding aberrations specific to a single gene.}
For every coding variant $i$ observed in at least one patient within a given gene in the exome data, there is an indicator variable $X_i$ capturing whether that variant affects the function of the protein.
The value of $X_i$  is inferred by our model. We use harmfulness predictions (as provided) as priors over the $X_i$ variables. 

For any considered individual and gene, there are hidden variables $Y_i=X_i \cdot g_i$,  where the genotype input $g_i$ for the $ i^{th}$ variant  is encoded as 0, 1 or 2 (indicating respectively Homozygous Reference, Heterozygous, Homozygous Alternative). $Y_i$'s  indicate whether the $i^{th}$ variant is functional in the given individual and whether it is in a homozygous or heterozygous state. 

To aggregate all functional variants affecting the same gene in the same individual, we define the quality aberration variable $Q \in \{0, 0.5, 1\}$. Its values respectively indicate a normally functioning, a  partially dysfunctional (e.g. haploinsufficiency) and a fully dysfunctional/inactivated protein.
The probability distribution of  $Q$ depends on the state of the $ \textbf{Y}$ and the relationship is described by the factor $\varphi_1(Q,  \textbf{Y}) = P(Q\mid \textbf{Y})$. For more details, see Supplementary Table 1.

\paragraph{Modeling gene expression aberration specific to a gene.}

The gene expression data is pre-processed by applying quantile normalization and correcting for batch effects if necessary. We then use $log(a+x)$ to transform the data, where $a\succ 0$ is a constant added to avoid the amplification of the variance in low-expression genes. Further, we apply robust standardization to each gene. The usual standardization of gene expression is done by subtracting the mean and then dividing by the standard deviation. The robust standardization uses measures that are less sensitive to outliers. 
If $e_0$ is the vector of log-transformed expression levels for a particular gene then we have:
\begin{equation}
e =\frac{e_0 -  median(e_0)}{K \cdot median(|e_0 -  median(e_0)|)}
\end{equation}
$K=\frac{1}{\phi^{-1} (\frac{3}{4})}\approx 1.4826$ is a multiplicative constant that makes the median absolute deviation (MAD) a consistent estimator of the standard deviation \cite{Leys2013}. The robust standardization is applied separately on cases and controls in order to remove the confounding effect of the phenotype (downstream consequences of having the phenotype on gene expression: such as drugs, immune response, etc.). The standardized levels of expression are represented by $E$ and are provided by the user as input to our method (see Figure \ref{fig:11}).

Variable $S$ is a categorical variable indicating which interval of expression levels $(e)$ can have an effect on the phenotype. $S$ has 5 categories: no expression level is aberrant, only the very low levels are abberant, only the low levels are abberant, only the high levels are abberant, only the very high levels are aberrant.
Since all values are the result of robust standardization, the intervals can be defined independently of the genes. For example, $e \leq -3$ defines the "only very low levels" interval  and $e \geq 2$ defines the "only high levels" interval. A uniform prior is used in the absence of the prior knowledge on which levels of expression are more likely to generate a functional disruption.

Similarly to $Q$, $A$ is a variable indicating mRNA expression level (Quantity) aberration. We have $A \in \{0, 0.5, 1\}$ (standing for no aberration, mild dysregulation, strong dysregulation respectively). 
The distribution of $A$ depends on the distribution of the $S$ variable and on the expression level in the gene and individual considered (See description of $\varphi_2$ factor in the Supplementary Table 2).

Finally, the quality aberration $Q$ from the exome data and the quantity aberration $A$ from the expression data are combined in an additive fashion to estimate the total functional disruption incurred by the studied gene in any given individual. The graphical representation of a single gene model is shown on the right panel of Figure \ref{fig:11}.

\paragraph{Whole genome model.}

The main goal of the model is to distinguish genes that are relevant to the disease from those that are not. For a given gene $i$, $H_i$ is equal to $1$ if the gene is relevant to the phenotype and $0$ otherwise. There are two types of priors used for the $H$ variables:
\begin{enumerate}
\item The gene network prior is encoded in the factors $\Theta$ connecting each pair of genes known to interact from existing protein-protein interaction networks. These factors make two interacting genes more likely to  both be relevant to a given phenotype than a random pair of genes that do not interact. It also helps to bring interacting genes into the set of phenotypically relevant genes if one or both of them are not marginally significant. We use asymmetric directed factors parametrized by the degree of the target gene in the network
to account for the large discrepancy in degree distributions for different genes (see Suppl. Materials 1.3 ).
\item The sparsity prior is encoded in two ways: a) a uniform small prior over all genes indicating that it is unlikely for any random gene to be involved in the phenotype; b) a regularization factor $\rho$ encouraging situations where only a few $H_i$ are active at once. 
\end{enumerate}

For every studied individual, there is one $G$ variable per gene indicating whether that gene has a functional importance in that particular individual. G is binary and depends on $Q$, $A$ and $H$ as described by the factor:
$
\varphi_3(G=1,H,Q,A)=P(G=1 \mid H,Q,A)= H\cdot max(1,A+Q)
$. 
Thus, a gene can only be relevant in a particular individual if the gene is shown to be relevant to the phenotype in general and if that particular individual has suspected aberrations in his exome variants or in his level of expression for that gene.

Finally, the phenotype (disease/no disease) is given as an input. It relates to the number of active $G$ variables in each patient: a patient is likely to have some affected genes while a healthy individual should have few or no affected genes.
The expected number of affected genes for patients and the tolerated number of affected genes in healthy population depend on the disease and its complexity and are therefore taken as parameters $\varphi_4$ by our method. (More on $\varphi_4$ in Supplementary equation (2)).

\paragraph{Inference}

We use a loopy belief propagation to jointly infer the marginal distributions of the unobserved variables in our graphical model modified for efficiency. All computed messages are normalized and kept in log space for numeric stability. We used message damping with parameter  $\alpha=0.4$ by default to improve the convergence behaviour of the algorithm.

Variables related to the aberrations in each gene and patient are the most numerous, and their messages are the most expensive to compute and the slowest to vary. Conversely, the messages between the phenotype, the $G$ and $H$ variables are the fastest to change since these variables are tightly correlated.
Therefore, for every time we update the messages for $X$, $Y$, $Q$  and $A$ variables, we update the messages between $G$, $H$ and phenotype variables multiple times or until their local convergence.

Some of the factors connect a large number of nodes to one node. For example: the $\varphi_4$ factor connecting $G$ variables to phenotype,  the regularization factor $\rho$ connecting the $H$ variables together, and the $\varphi_4$ factor connecting $Y$ nodes to a $Q$ node.
Since the factor mainly depends on the sum of the variables in each of these examples, the computation becomes possible by estimating the distribution of the sum of variables and then using it as an intermediate variable. We use the poisson binomial to estimate the sum of bernoulli variables. 
In a recent review of Poisson Binomial estimation methods  \cite{Hong2013}, the complexity of the methods presented is at least O($n^2$), which is inefficient for messages computation especially for large number of variables $n$. Here we develop a divide and conquer approach for computing the poisson binomial in O($n log^2(n)$).
This was done based on the simple observation that the sum of two Poisson binomial variables $Z_1$ an $Z_2$ is a poisson binomial variable Z and the distribution of Z is the convolution of the two vectors representing the distributions of $Z_1$ and $Z_2$ (by analogy to a product of two polynomials).
Messages to/from cardinality potentials were previously computed in the same complexity using a very similar approach \cite{Tarlow2012}, but it was not previously formulated as a solution for estimating poisson binomial distributions.

After convergence of the loopy belief propagation our method returns the posterior marginal probabilities for the $H$ and $G$ variables, along with the list of most relevant abberrations (coding variants or expression levels) in each affected individual according to the model.

\section{Results}

Since there are very few complex diseases with known gene abberations, but there are very good genome simulators of disease, particularly of rare variants, we first analyze our performance using an extensive set of simulations. We then showcase our method on a real breast cancer dataset. But first, we list the sources of our biological priors.

	Harmfulness predictions are used as priors over the $X_i$ variables indicating whether the coding variant $i$ affects protein function. These predictions are obtained using public methods and libraries. We used ANNOVAR \cite{Wang2010} for variant annotation. We then used Polyphen2 scores as priors over the nonsynonymous variants. Variants annotated as nonsense or loss-of-function (LoF) are systematically assigned a high prior (such as 0.99). For now, we ignored variants that are not annotated as nonsynonymous or nonsense.
The use of harmfulness predictions is optional. Since they are only used as priors, their absence does not usually have a large impact on the final result (see below).

 We used the BioGrid human protein-protein interaction network \cite{Chatr-Aryamontri2015} as a prior for gene connectivity. 

\subsection{Simulations}

\paragraph{Overview of the simulation process.}

For simulating coding variants, we used the European population model with the optimal parameters as in \cite{kryukov2009power}. We considered the single nucleotide variants (SNVs) with selection coefficient greater than s=0.001 as the deleterious SNVs and the remaining SNVs as neutral, as recommended in \cite{price2010pooled}. We then sampled harmfulness scores from real Polyphen scores distribution computed on known deleterious variants and neutral variants as in \cite{price2010pooled}. The generated missense SNVs were partitioned among the 6900 genes present in both BioGrid annotation and in the healthy gene expression dataset we used as background for the distribution of genes' levels. For every individual and gene, we randomly decided with rate $\lambda_0$ whether there will be a perturbation of expression level in that individual for the considered gene.

The procedure above was done on a population level. To simulate a disease scenario, we selected a number of genes that are close to each other on the protein-protein interaction network as being the causal disease mechanism. For every individual, we count the number of harmful mutations and the number of expression perturbations within the disease mechanisms and we considered that sum as indicative of the phenotype.
The disease cases were then selected to be the top 3000 individuals according to that sum (we suppose the disease prevalence in the population is 0.3\%).
 
For the expression data, we started from a healthy expression dataset (GEO: GSE9006)  as a background signal of expression. For every individual selected (patient or control), we generate gene expression levels by sampling with noise from the real data and then applying the gene expression perturbations simulated earlier on. The magnitude of the change in expression was uniformly sampled from $[2\sigma, 4\sigma]$ where $\sigma$ is the standard deviation for the considered gene expression (background signal). The sign of the expression changes is chosen randomly for each gene.

We varied several simulation parameters to check robustness of our method. First, we varied the sample size: we selected $n$ patients from 3000 effected individuals and $n$ controls from the rest of the population. Second, we varied the number of causal genes. Finally, we varied the proportion $\lambda$ of aberrations related to expression versus coding variants. 

We characterize the performance of our  method by how well the disease mechanism was reconstructed (sensitivity, precision, N-Precision) and compare our results to those of four rare variants aggregation methods: CAST \cite{hoffmann2010comprehensive}, C-ALPHA\cite{neale2011testing}, SKAT \cite{Wu2011} and RWAS\cite{sul2011optimal} along with differential expression and the dmGWAS method \cite{Jia2011} that uses networks in gene detection. From the output of our method, we consider a gene as a positive (disease associated) if its marginal probability is larger than $0.5$. For all the other methods except dmGWAS, gene p-values were used as the criterion for ranking and calling genes. A gene was considered a positive if its associated p-value was lower than the significance threshold ($< 8.10^{-6}$ after multiple hypothesis correction). dmGWAS returns fixed lists of genes. We use the highest scoring list as the positive gene calls.

\subsubsection{Performance}

\begin{center}
\begin{figure}[!h]
\centerline{\includegraphics[scale=0.35]{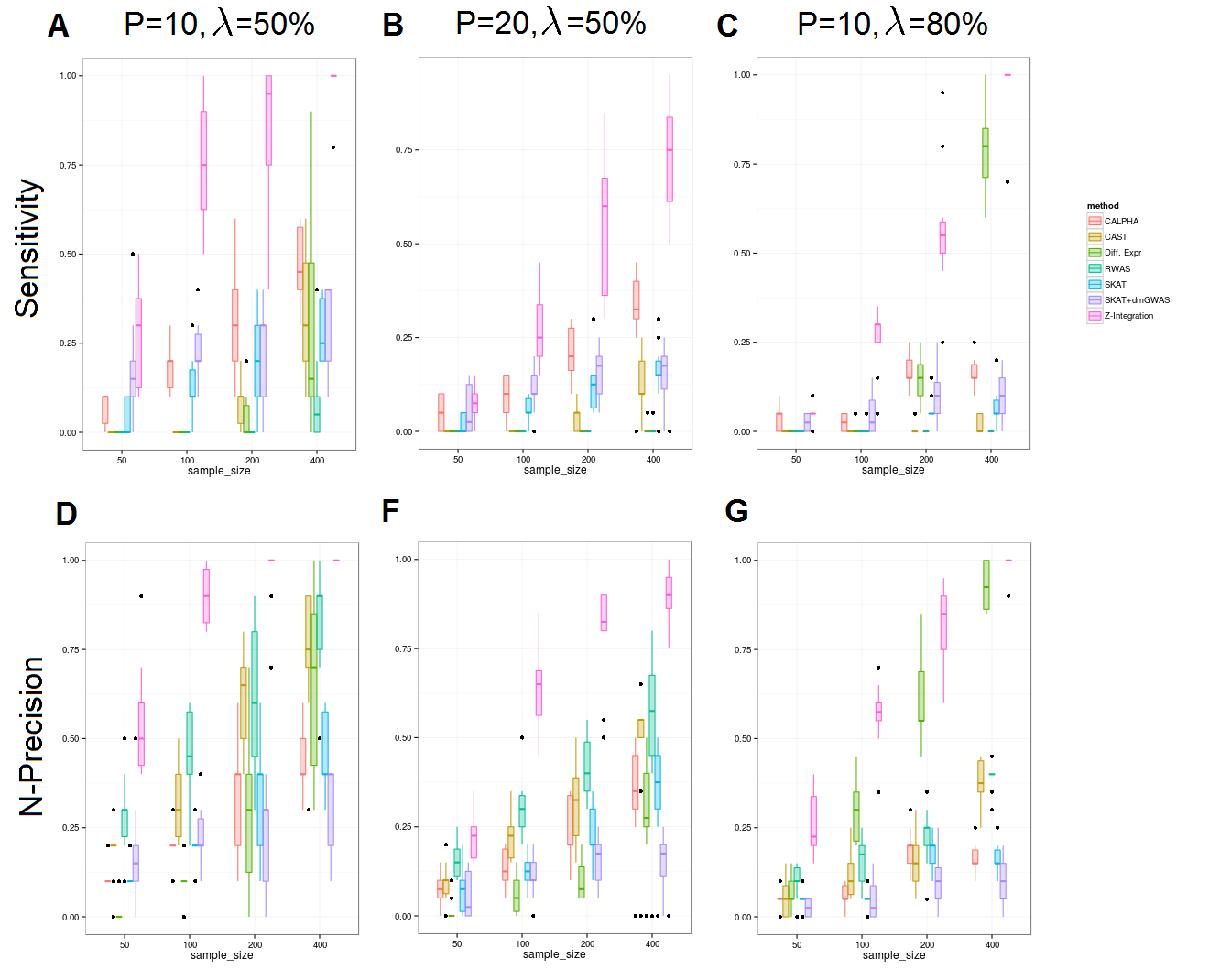}}
\caption{Sensitivity and N-Precision as a function of the simulation parameters, evaluated for our method (in pink) and compared to other variants' aggregation methods and to differential expression. The sample size is varied from $50$ to $400$ on the x-axis of each single graph (The sample size is equal to the number of cases, which is equal to the number of controls). The effect of changing $P$ and $\lambda$ is shown in the different columns. $P$ is the number of genes simulated to be causal and $\lambda$ is the proportion of simulated aberrations that are regulatory (expression) as opposed to coding (Exome). Sensitivity is the proportion of simulated causal genes that are detected by the method. N-Precision  is computed based on the top N=P genes reported by each method and is equal to the proportion of true causal genes among those. }
\label{fig:01}
\end{figure}
\end{center}

Figure \ref{fig:01}-A,B,C shows that our method has higher sensitivity to the causal genes than any other method considered and that it is often the only method capable of detecting the disease mechanism in many realistic scenarios. In particular, the N-Precision measure (how likely is the true gene to be ranked within the top N genes) in Figure \ref{fig:01}-D,E,F shows that it is not a problem of stringent p-value threshold for the other methods but that our performance is higher for the task of prioritizing candidate genes in general and that the causal genes are systematically ranked higher by our method. 
For the criterion used to determine positive gene calls (marginals $\geq 0.5$), the precision of our method was equal to 1. It is not the case for the other methods, some of which suffered from large type 1 errors.

As expected, we see that the problem of identifying the causal genes is more difficult for smaller sample sizes and for more complex disease (larger $P$ in Figure \ref{fig:01}).
We can see that the performance of most methods drops considerably when the contribution of the coding variation is only $20\%$  (i.e. aberrations in expression levels constitute $80\%$ of the causes) instead of $50\%$. This is particularly interesting because the ratio $20\%$ to $80\%$ for respectively coding and regulatory aberrations is a realistic scenario as mentioned in the literature \cite{adeyemo2009genetic}.

\subsubsection{Robustness}
\label{robust}

\begin{figure}[!htpb]
\centerline{\includegraphics[scale=0.4]{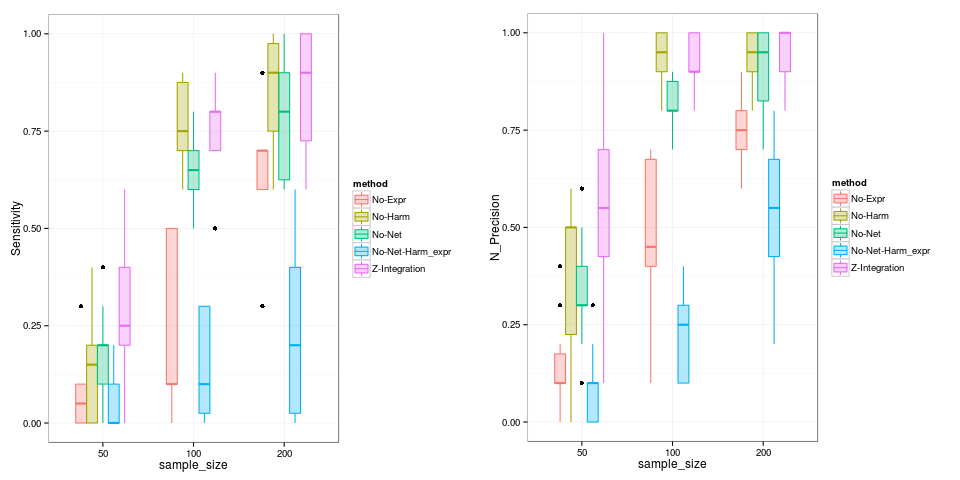}}
\caption{Sensitivity and N-Precision as a function of the sample size used and comparison between multiple runs of our method using different subsets of the inputs. The goal is to see the contribution of different input types to the performance. We fixed $P=10$  and the Expression versus SNPS causality ($\lambda$) to 50\% (Both have equal contribution to the phenotype on average). Each bar describes the results of 10 simulations. }
\label{fig:05}
\end{figure}

Our method takes multiple biological sources of information and annotations as inputs in order to enhance the power to detect genes relevant to the phenotype. Figure  \ref{fig:05} shows how much each of these additional inputs contribute to the power of the method. Specifically, it shows that running our method without using the gene network or without harmfulness predictions results in only a small drop in performance. However, running our method without expression data, does have a noticeable impact. This is expected since half the disease signal is contained in the expression data in this particular set of simulations. Still, even running without expression our method largely outperforms all others approaches as shown by comparing Figure \ref{fig:05} and Figure \ref{fig:01}-A,D. 
It is only when we use our method without any of these additional inputs that we see a substantial drop in performance, making the performance of our method on par with the other methods.

We also checked that our method is still able to reconstruct the disease mechanism when there is a large discrepancy between the contributions of the causal genes (see Supplementary Figure 9) and is robust to changes in regularization parameter (Supplementary Figure 10).

\subsection{Breast cancer data}

We used exome sequencing and gene expression data (RNA-Seq) for 826 breast cancer patients from 4 subtypes (238 Luminal A, 359 Luminal B, 148 Basal, 81 HER2+). All data was downloaded from The Cancer Genome Atlas (TCGA). The data for 20,000 genes was available and was used in the method directly, without gene pre-selection.
We analysed all pairs of breast cancer subtypes (using the first subtype in each pair as patients and the second as controls and vice versa) and report the genes with marginals $>0.5$. We balance the number of patients and controls by subsampling from the subtype with the larger sample size.
For the luminal subtypes the recurrent genes were PIK3CA, CDH1 and GATA3, with luminal A patients having more often mutations in CDH1, MAP3K1 and MLL3, while luminal B patients having more often TP53 and GATA3 mutations when we compare both subtypes to each other.
The HER2 subtype is the one with the smallest sample size and the only gene with marginal $>0.5$ is TP53, which appears when we compare HER2 to Luminal A or Luminal B subtypes.
Finally, the basal subtype is characterized by more TP53 aberrations than any other subtype. The TTN gene is also associated with this subtype. 

Every one of the genes reported by our analysis was already identified by at least one previous study as characteristic of the corresponding subtype of breast cancer \cite{Greenman2007,Wang2011,Ellis2012}
Some of the genes reported by our analysis were supported by aberrations in the exome of some patients, and abnormal levels of expression in other patients. This means that by integrating both data types we gained power to detect those genes.

\section{Discussion}

In this paper we propose an integrative graphical model framework for combining exome and gene expression data to increase the power of detecting genes associated with complex human diseases. The main contribution of the paper is the principled approach to combining these two data sources in a biologically sensible way. While people have used genetic variants to predict gene expression (eQTL analysis), combining this data as two complementary sources has not been done to date. We made sure to incorporate additional biological information, such as variant harmfulness and gene interactions as priors. This again is an innovation, since gene networks have been mostly used to either pre-select or combine the identified variants, whereas we propose to use it as part of the inference process. Additionally, using networks as priors, rather than in the post-hoc analysis, allows the method to be less sensitive to the inherent noise present in the network. 

We demonstrated that even given a small sample size, our method is able to outperform several other widely used methods (using either gene expression or exome alone). Our simulations show that the method is not too sensitive to any of the priors and that the combination of this information gives a great boost to the gene identification. The breast cancer analysis showed that we can recover known breast cancer genes.

We have to note that it is important to take great care when using gene expression data in cancer. Expression data in cancer represents the downstream effect of many genes through their interactions. It is not unusual to observe a significant gene expression change in thousands of genes, the majority being a downstream, rather than the driver, effect  (e.g. inflammation, drug response, etc) Additionally, and more importantly, there is a large heterogeneity in gene expression in cancer: many patients within the same subtype will appear to have an abberant expression. These variations are of unknown cause. It is desirable to remove the broad variability patterns while preserving consistent changes across a few patients. In our work, we used the quantile normalization to address this point, but we will be exploring this question further to establish a cancer specific pre-processing framework.

We will also further investigate the translational impact of this work, analyzing the posterior over genes for each patient individually. This will help to assess the heterogeneity in cancer in general, but more so, to potentially assess the ability of each patient to respond to mutation targetting drugs.

\bibliographystyle{plain}
\bibliography{diseasemechanismshort}

\end{document}


\maketitle

\section{Methods}

\subsection{Graphical model}

\begin{figure}[!h]
\centerline{\includegraphics[scale=0.4]{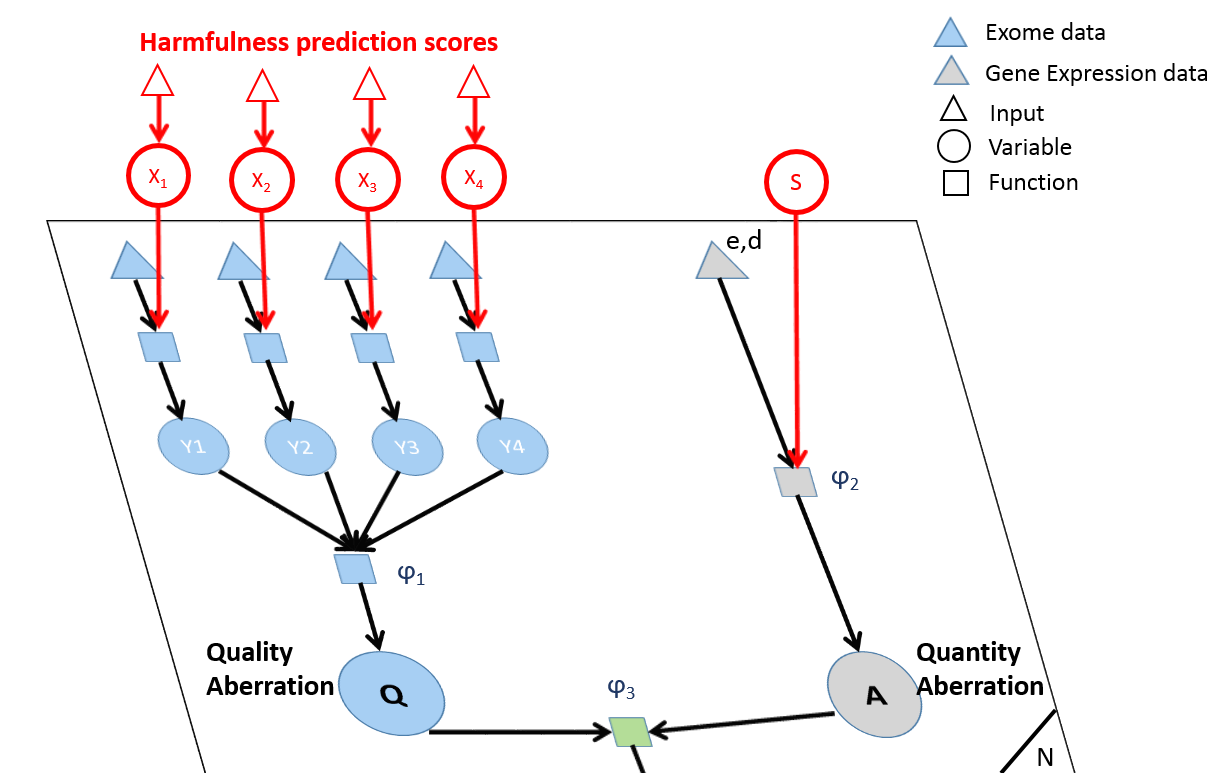}}
\caption{Upper part of the graphical model that describes one gene. All the variables, factors and inputs inside the plane are individual-specific and the whole plane should be replicated for every individual. The variables, factors and inputs outside the plane are not individual-specific. }
\label{fig:11}
\end{figure}

\begin{figure}[!h]
\centerline{\includegraphics[scale=0.5]{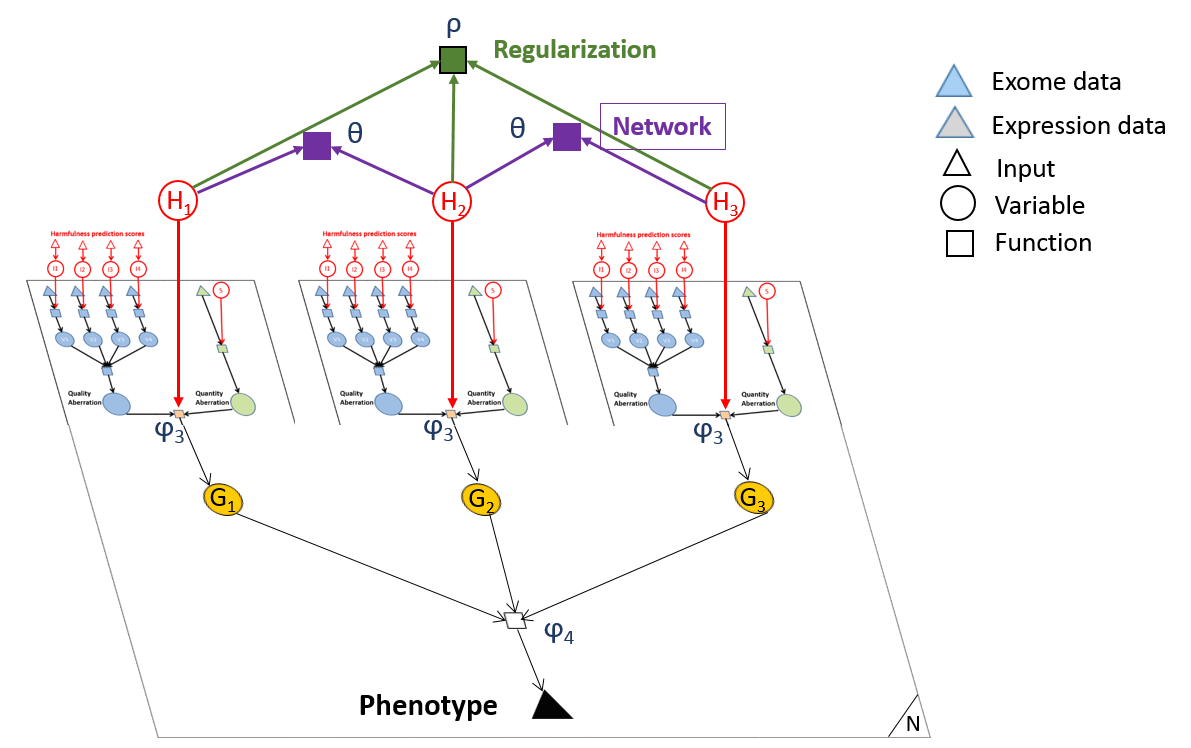}}
\caption{Whole graphical model of the genotype-phenotype association illustrated with only 3 genes. All the variables, factors and inputs inside the plane are individual-specific and the whole plane should be replicated for every individual. The variables, factors and inputs outside the plane are not individual-specific. }
\label{fig:12}
\end{figure}

The probability distribution of  $Q$ depends on the state of the $ \textbf{Y}$ variables and the relationship is described by the factor $\varphi_1(Q,  \textbf{Y}) = P(Q\mid \textbf{Y})$. In our implementation we use the  $\varphi_1$ factor  in table \ref{tab:Q}:

\begin{table}[h]
\centering
\begin{tabular}{ c || c | c | c } \hline 
$Q|\textbf{Y}$ & $\exists i$ such that $Y_i=2$ &  $\exists i$ such that $Y_i=1$ , and $\forall i$ $Y_i<2$   & $Y_i=0$ , $\forall i$ \\ \hline                        
0 & $0$ & $0$ & $1$ \\
0.5 & $0$ & $ 0.5^{(-1+\sum{Y}) }$ & $0$ \\
1 & $1$ & $1- 0.5^{(-1+\sum{Y}) }$ & $0$ \\
\hline  
\end{tabular}
\caption{$\varphi_1$ factor or the conditional distrubution of $Q$ given $Y$}
\label{tab:Q}
 \end{table}

In other words, if the individual have no functional coding variants at all in the considered gene (i.e.  $Y_i=0$ , $\forall i$), we have  Q is necessarily 0 (No quality aberration in the considered gene and individual).
If there is an $i$ such that $Y_i=2$ (i.e. the individual have at least one homozygous functional variant in the considered gene), then we have Q is necessarily 1 (both alleles are compromised).
In the absence of homozygous functional variants, the probability distribution of $Q$ also depends on the number of heterozygous variants and their repartition across alleles. We assume the haplotypes are not phased.
In that case, if there is an  $i$ such that $Y_i=1$ (i.e. the individual have heterozygous functional variants in the considered gene), we have the probability of $Q=0.5$ (the gene is only partially dysfunctional) is the probability of all variants falling into the same allele. Therefore,
\begin{equation}
\varphi_1(Q=0.5,\textbf{Y})= P(Q=0.5\mid \textbf{Y})= 0.5^{(-1+\sum{Y}) } 
\end{equation}

 More specifically the relationship between the variables is described by the $\varphi_2$ factor.
In our implementation, we chose the $\varphi_2$ factor defined by the table \ref{tab:A} 

\begin{table}[h]
\centering
\begin{tabular}{ c || c | c } \hline 
$A|e,S$ & $e \not\in S$ &  $e \in S$  \\ \hline                        
0 & $1$ & $0$ \\
0.5 & $0$ & $ 1_{|E|< 3} + 0.5\cdot 1_{|E|\geq 3}$ \\
1 & $0$ & $ 0.5\cdot 1_{|E|\geq 3}$ \\ 
\hline 
\end{tabular}
\caption{$\varphi_2$ factor or the conditional distrubution of $A$ given $S$ and $e$}
\label{tab:A}
 \end{table}

In other words, as long as the expression level $e$ for the considered individual and gene is outside the interval defined by $S$, there is no regulatory aberrations ($A=0$). However if $e \in S$ then there is a regulatory aberrations and its severity depends on how extreme $e$ is.

As an extension to $\varphi_4$, we added an additional free parameter that control the number of active genes needed to generate the disease. $C$ stands for disease complexity and it parametrizes the relationship between the number of active genes per individual and the phenotype as follows: Affected individuals are encouraged to have at least $C$ active genes while unaffected individuals are expected to have less.
$C$ takes values in the low integers (can be 1 to 4 for example) and we can put a uniform prior on its distribution if we do not have a prior knowledge about the disease complexity. 

In our standard implementation, $C$ is fixed to $1$ (equivalent to having two heterozygous mutations in two important genes in every patient), the parameter $p_{fc}$ is given as input, and we have:
\begin{equation}
\begin{split}
\varphi_4(Phenotype=case, G,C=1) &= P(Phenotype=case \mid G,C=1)\\
                                                  &=
\begin{cases}
p_{fc}, & \text{if}\ \sum_i{G_i} =0 \\
1- (1-p_{fc})^{\sum_i{G_i} }, & \text{otherwise}
\end{cases}
\end{split}
\end{equation}

The parameter $p_{fc}$ is the probability of false case or more explicitly, it is the probability of an individual with no active disease associated genes to be a case rather than a control. This parameter has a default value of $0.95$ that can be changed by the user.
In all our subsequent results we kept the default value.

\subsection{Belief Propagation Algorithm}

\begin{figure}[!h]
\centerline{\includegraphics[scale=0.4]{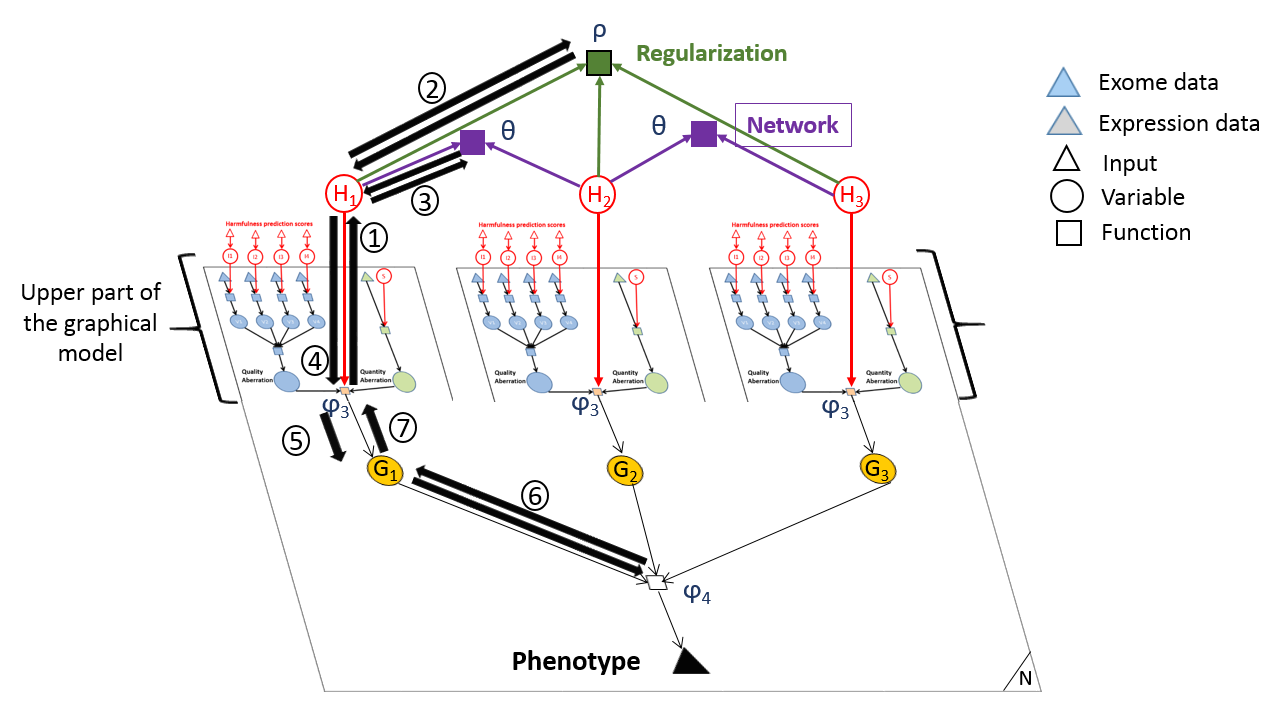}}
\caption{Order of message computations in the belief propagation algorithm. We start from the upper part of the graphical model and descend until we reach the double loop described by the numbers 1 to 7. The double loop is traversed a certain number of times or until local convergence and then we compute the backward messages for the upper part of the model. The whole process is repeated until convergence}
\label{fig:13}
\end{figure}

The formulas for the messages' computations are generated following the conventional approach defining two types of messages: factors to nodes and nodes to factors. 
\begin{figure}[!h]
\centerline{\includegraphics[scale=0.4]{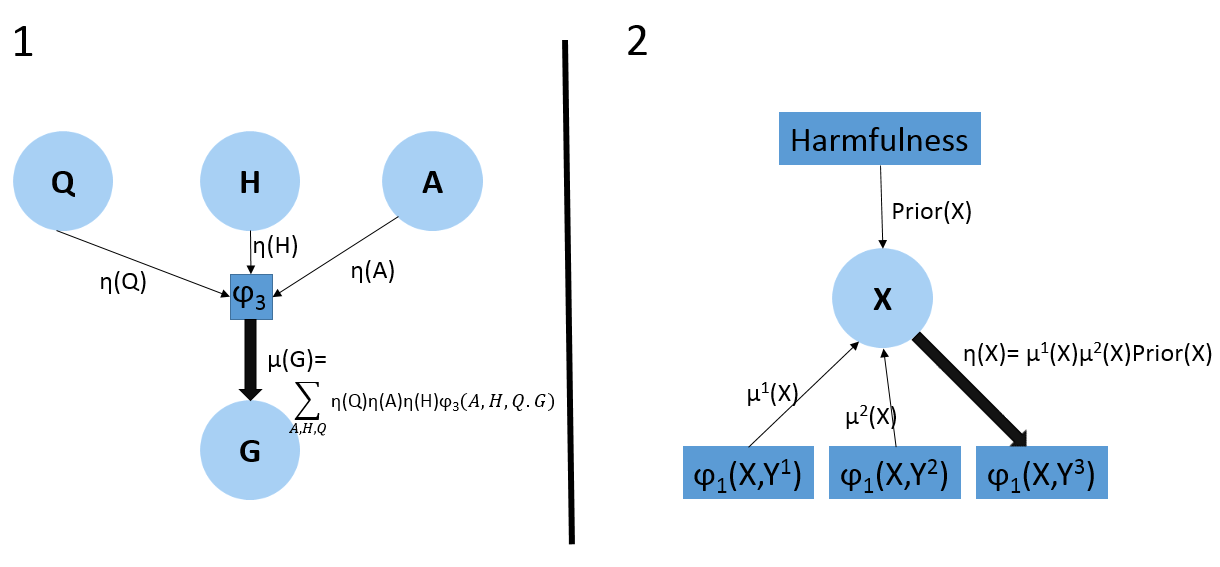}}
\caption{Examples of the types of messages computed in our algorithm. \textbf{1} factor to node message $\mu$ . \textbf{2} node to factor message $\eta$}
\label{fig:14}
\end{figure}
Figure \ref{fig:14} illustrates how these two types of messages are computed by using two examples from the graphical model. 
The message $\mu$ going from a factor to a node is the marginalization over all the other nodes of the product of all incoming messages $\eta$ by the factor itself. 
The message $\eta$ going from a node to a factor is equal to the product of all the incoming $\mu$ messages to that node from all other neighbour factors.
Two important remarks on Figure \ref{fig:14}-2 :
\begin{itemize}
\item The priors are also factors. For example, having a harmfulness prediction prior $h$ over a coding variant is the same as sending a message $(1-h,h)$ to the corresponding X variable.
\item The variable $X$ is not individual specific but is linked to the $Y$ variable from every individual, through the $\varphi_0$ factor. Therefore each $X$ variable is actually linked to $N$ $Y$ variables, where N is the number of individuals (On Figure \ref{fig:11}, every $Y$is on the plate and should therefore be replicated N times). 
\end{itemize}

The order chosen for updating the messages helped with the speed of the algorithm and the convergence. We further improved the convergence behaviour of the algorithm by damping messages which help avoiding oscillatory behaviours. Every time we update a message $\mu$ the new value becomes
\begin{equation}
\mu_{updated} = (1-\alpha) \mu_{old} + \alpha \mu_{new}
\end{equation}
, where $\alpha$ is the damping parameter (by default we take $\alpha=0.4$).

\subsection{Implementing gene network priors}

The gene network is used as prior over the $H$ variables indicating whether a gene is relevant to the disease.
It is encoded by the $\Theta$ facors which are placed between each pair of interacting genes.
$\Theta$ encourage pairs of interacting genes to be active to be active together for the same disease.
There are difficulties associated with the implementation of this prior:
\begin{itemize}
\item The connectivity degree is very different across gene. While some genes can be connected to tens or even hundreds of neighbours, other genes have very few or no connections. This can create large biases pushing some genes to be always spuriously associated with the phenotype or pushing genuinely important genes down.
\item We want to encourage interacting genes to be relevant together in the same disease, but we do not want to penalize genes when their neighbours are not related to the disease. An edge in the gene network means the two genes involved could potentially have similar function in relation to disease. The absnce of such an edge is not meaningful in itself.
\item Gene networks are considered a very incomplete annotation with probably a large proportion of missing edges. It is also possible that genes have inderect interactions with each other which makes them relevant to the same disease (second order neighbours).
\end{itemize}

To solve this problem and avoid having this prior biasing our results, we take the following measures:

\begin{itemize}
\item Instead of the symetric indirected factors usually used for modeling joint probabilities, we make the $\Theta$ facors into a double directed and asymetric factors. The idea is that we want to encourage a gene to be "on" if its neighbour have some evidence of being associated with the disease. But we do not want to encourage a gene to be "off" because its neighbours is. If the neighbor is off that should be noninformative of the state of the considered gene.
\item We parametrize the directed factor by the degree of the target gene. The goal is that the accumulation of small signals from a large number of non associated neighbours should never be enough to make a highly connected gene seem disease associated.
\item We fix an upper limit to the amount of contribution that can be received from all the active neighbours of a particular gene. Every gene starts from a small prior and need evidence from its own data (exome or expression) to reach large marginals. We make sure that the contribution of the network alone is never enough on its own to make the gene associated with the disease and that there must be enough examples of patients with aberrations in that gene for it to reach high marginal probabilities.

\end{itemize}

\subsection{Simulated data}

\subsubsection{Overview of the simulation process}

For simulating coding variants, we used the European population model with the optimal parameters as in \cite{kryukov2009power}. We generated missense SNVs (single nucleotide variants) for 900,000 individuals, each SNV is generated with a selection coefficient $s$. Overall there was around half a million SNVs most of which are very rare since the corresponding mutations happened in the recent population expansion period.
We considered the SNVs with selection coefficient greater than s=0.001 as the deleterious SNVs and the remaining SNVs as neutral, as recommended in \cite{price2010pooled}. We then simulated Polyphen2 scores for all SNVs. These scores are sampled from real Polyphen scores distributions computed on known deleterious variants and neutral variants in \cite{price2010pooled}.(We also tried Gaussians with mean respectively equal 0.2 and 0.8 for neutral and deleterious SNVs, with standard deviations equal 0.4).
The generated missense SNVs were partitioned among the 6900 genes present in both BioGrid annotation and in the healthy gene expression dataset we used as background for the distribution of genes' levels. At first, we treated all genes as if they were of equal length, i.e. the expected number of SNVs is the same across genes. Later on, we repeated our experiments for when the ratios of genes' lengths are sampled from an actual annotation of the lengths of the coding regions of all human genes (See Figure TODO supplements  fig:04). 
For every individual and gene, we randomly decided with rate $\lambda_0$ whether there will be a perturbation of expression level in that individual for the considered gene.
At this point, we have a list of all SNVs and expression perturbations for every individual and in each gene.

\begin{figure}[!h]
\centerline{\includegraphics[scale=0.4]{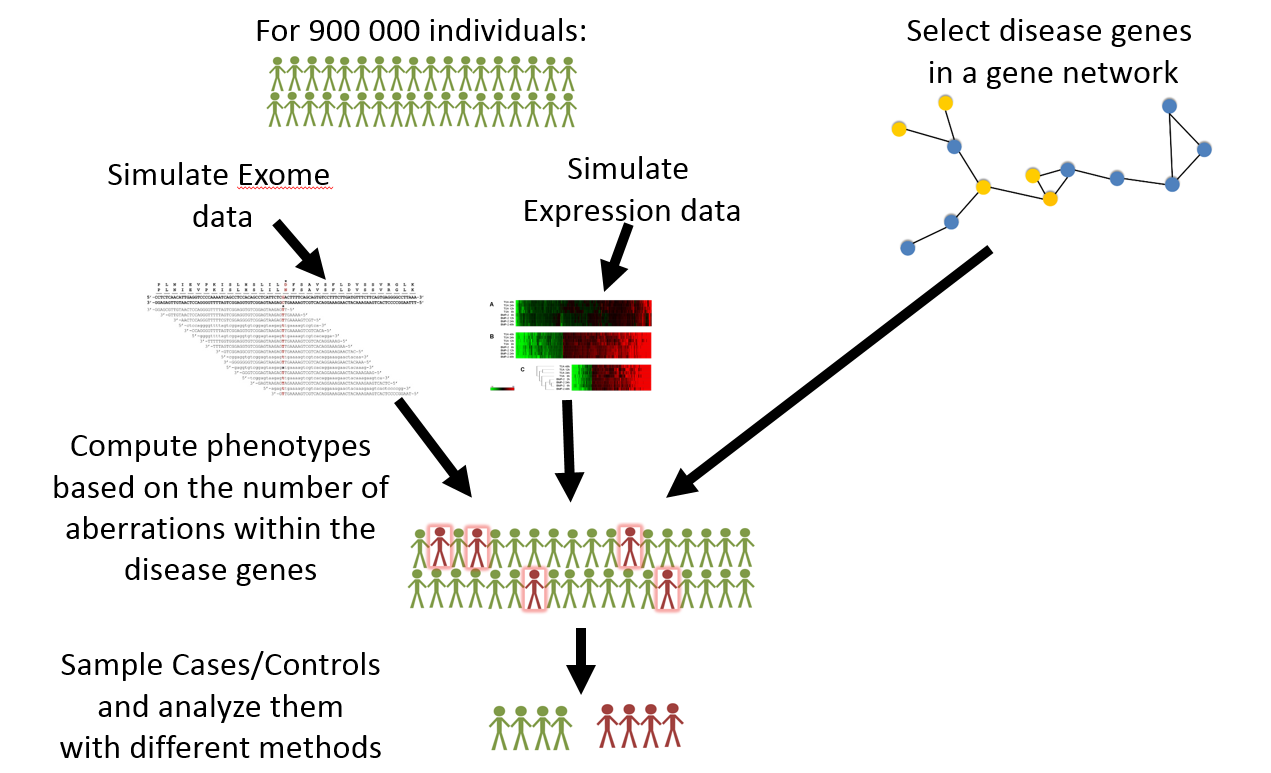}}
\caption{Framework of the simulations }\label{fig:15}
\end{figure}

\subsubsection{Performance}

\begin{figure}[!h]
\centerline{\includegraphics[scale=0.3]{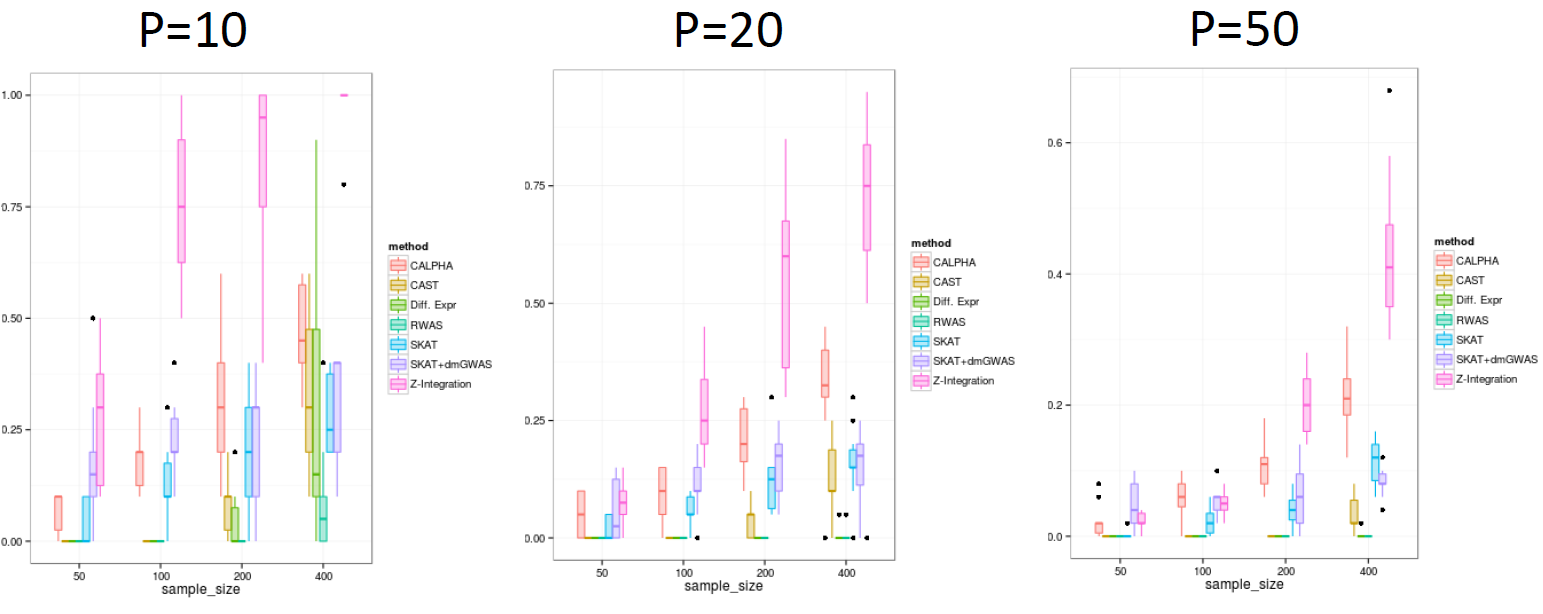}}
\caption{Sensitivity as a function of the sample size used and comparison with the results of variants aggregation methods and differential expression. For the three graphs, the simulated model contains respectively P=10, 20 and 50 causal genes and we fixed Expression-SNPS causality to 50\% (Both have equal contribution to the phenotype on average). Each bar describes the results of 10 simulations. }\label{fig:01}
\end{figure}

\begin{figure}[!h]
\centerline{\includegraphics[scale=0.3]{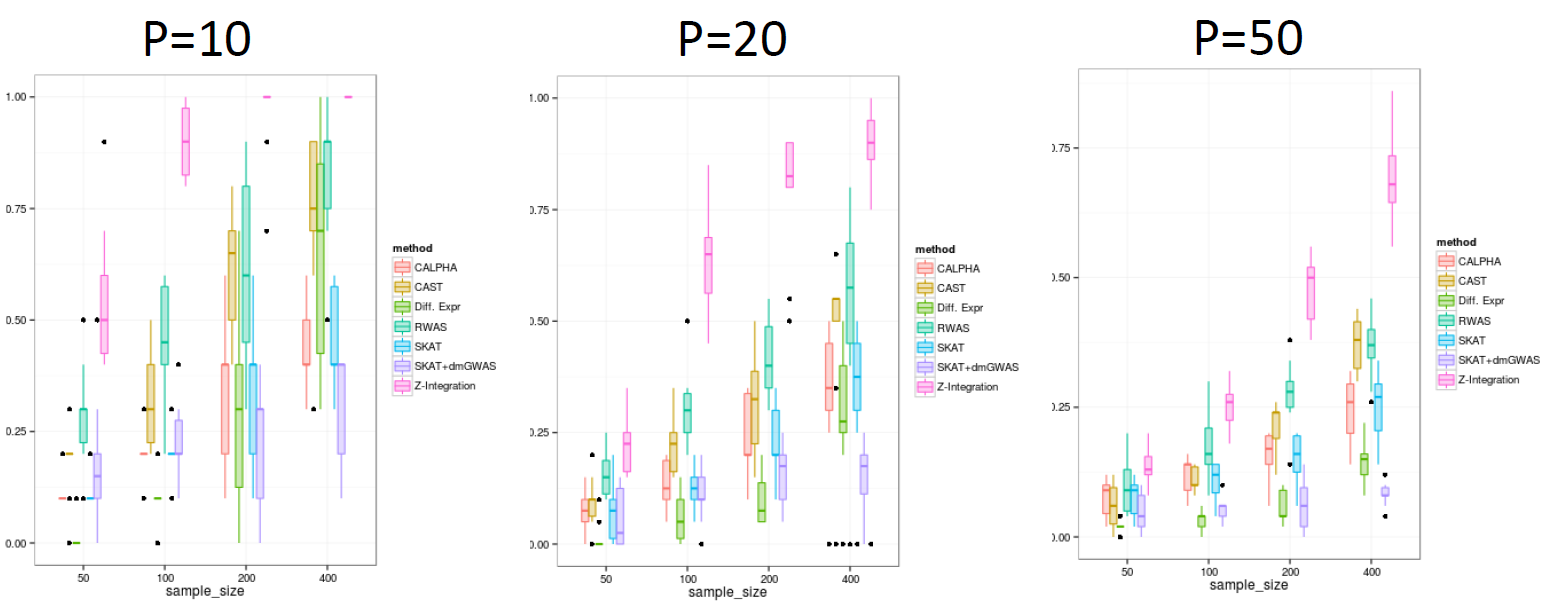}}
\caption{N-Precision as a function of the sample size used and comparison with the results of variants aggregation methods and differential expression. For the three graphs, the simulated model contains respectively P=10, 20 and 50 causal genes  and we fixed Expression-SNPS causality to 50\% (Both have equal contribution to the phenotype on average). Each bar describes the results of 10 simulations. N-Precision is computed based on the top N=P genes reported by each method. }\label{fig:02}
\end{figure}

\begin{figure}[!h]
\centerline{\includegraphics[scale=0.5]{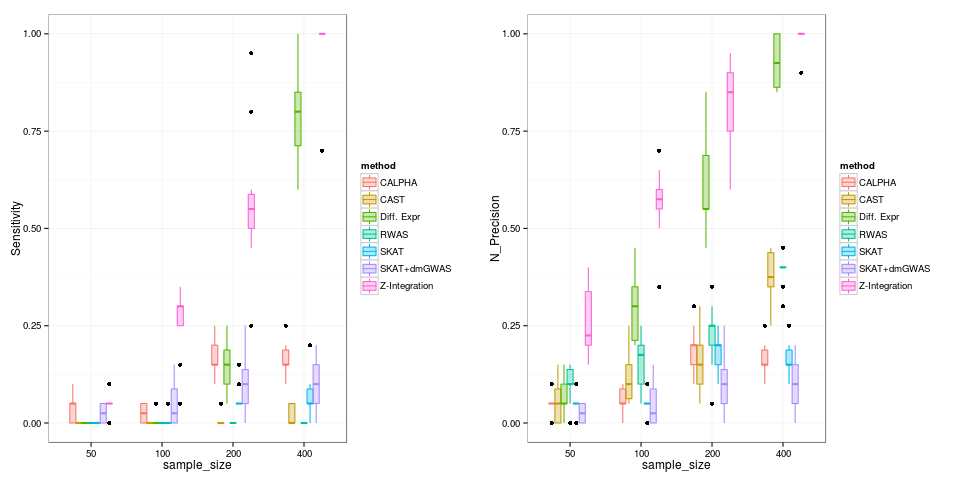}}
\caption{Sensitivity and N-Precision as a function of the sample size used and comparison with the results of variants aggregation methods and differential expression. The simulated model contains  P=20 causal genes  and we fixed Expression-SNPS causality to 80\%. Each bar describes the results of 10 simulations. N-Precision is computed based on the top N=P genes reported by each method. }\label{fig:03}
\end{figure}

\begin{figure}[!h]
\centerline{\includegraphics[scale=0.5]{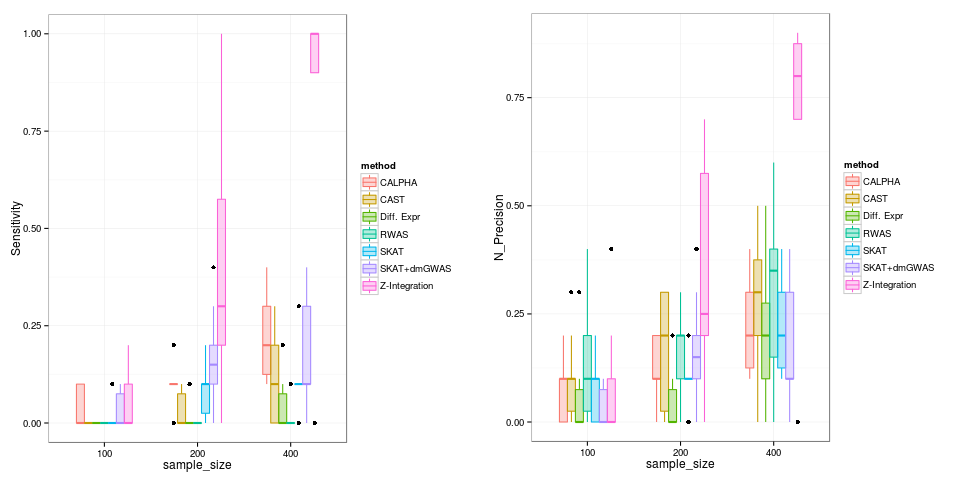}}
\caption{Sensitivity and N-Precision as a function of the sample size used and comparison with the results of variants aggregation methods and differential expression. The simulated model contains  P=10 causal genes  of different lengths (sampled from real human coding regions' lengths distribution) and we fixed Expression-SNPS causality to 50\%. Each bar describes the results of 10 simulations. N-Precision is computed based on the top N=P genes reported by each method. }\label{fig:04}
\end{figure}

Finally the problem is made more difficult when the causal genes have highly discrepant marginal contributions to the disease. In Figure \ref{fig:04}, we assign different lengths to genes, sampled from real coding regions lengths. In consequence, longer genes are mutated much more often than others and we end up with very few genes with strong signal and a majority of causal genes that are very hard to detect because they harbor very few coding variants across affected individuals. In this type of scenario (a few genes with strong signal and many with very weak signal), our method is the only one that is still able to recover most of the causal genes because it uses the gene network which help strengthen the signal of neighbour genes and because it uses the expression levels aberrations which frequencies is not affected by the length of the genes' coding regions in our simulations.

\subsubsection{Robustness}

\begin{figure}[!h]
\centerline{\includegraphics[scale=0.5]{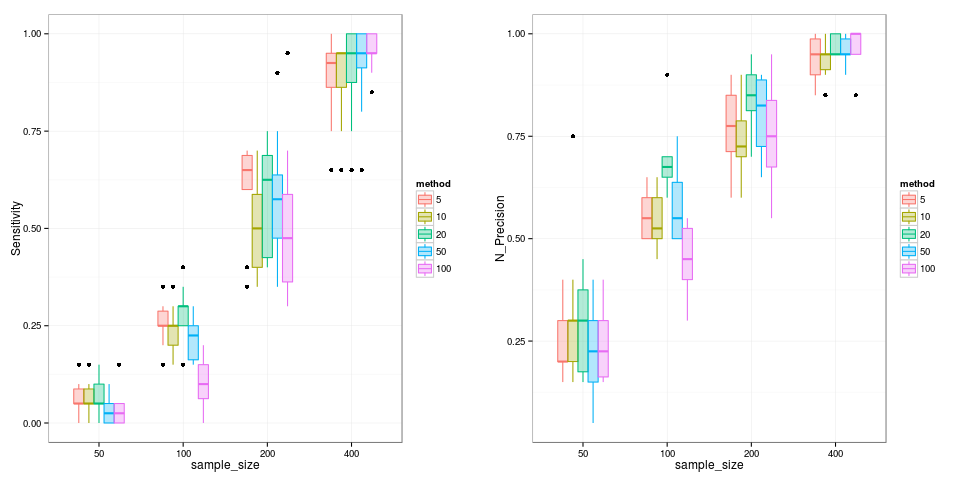}}
\caption{Sensitivity and N-Precision as a function of the sample size used and comparison between multiple runs of our method using different values for the $d$ parameter. The goal is to see the robustness to the choice of $d$. We fixed P=20  and the Expression-SNPS causality to 50\% (Both have equal contribution to the phenotype on average). Each bar describes the results of 10 simulations. }\label{fig:06}
\end{figure}

Our method also use as input the parameter $d$ indicating the expected number of causal genes. This parameter is used for regularization purposes to penalize solutions involving too many genes and encourage solutions that have close to $d$ genes.
Figure \ref{fig:06} shows that the result of our method is very robust to the choice of $d$ parameter. As long as $d$ is within an order of magnitude from the true number of causal genes, the exact choice of $d$ does not affect the sensitivity, the precision or the N-Precision.
For example, in analyzing a disease that have 20 causal genes, using $d=10$ or $d=50$ instead of $d=20$ does not affect the number of correctly detected genes and does not substantially add to the false positives.

\bibliographystyle{plain}
\bibliography{diseasemechanismshort}